\documentstyle[11pt,aaspp4,epsfig]{article}
\begin{document}
\title{Simultaneous ASCA And RXTE Observations of Cygnus X--1 During Its 1996 State Transition}
\author{Wei Cui\altaffilmark{1}, K.~Ebisawa\altaffilmark{2}$^,$\altaffilmark{3}, T.~Dotani\altaffilmark{4}, and A.~Kubota\altaffilmark{5}}
\altaffiltext{1}{Center for Space Research, Massachusetts Institute of Technology, Cambridge, MA 02139}
\altaffiltext{2}{code 660.2, NASA/GSFC, Greenbelt,MD 20771}
\altaffiltext{3}{also Universities Space Research Association}
\altaffiltext{4}{Institute of Space and Astronautical Science, Yoshinodai, Sagamihara, Kanagawa, 229, Japan}
\altaffiltext{5}{Department Physics, University of Tokyo, Hongo, Bunkyo-ku, Tokyo, 113 Japan}
\authoremail{cui@space.mit.edu, ebisawa@subaru.gsfc.nasa.gov, dotani@astro.isas.ac.jp, aya@miranda.phys.s.u-tokyo.ac.jp}
\slugcomment{Revised and re-submitted to the {\it Astrophysical Journal Letters} on 10/31/97}
\begin{abstract}

We report results from simultaneous ASCA and RXTE observations of Cygnus~X--1 
when the source made a rare transition from the hard (= low) state to the soft 
(= high) state in 1996. These observations together cover a broad energy range 
$\sim$ 0.7--50 keV with a moderate energy resolution at the iron K-band, thus
make it possible to disentangle various spectral components. The low-energy 
spectrum is dominated by an ultra-soft component, which is likely to be the 
emission from the hottest inner portion of the accretion disk around the black
hole. At high energies, the X-ray spectrum can be described by a Comptonized 
spectrum with a reflection component. The Compton corona, which upscatters soft
``seed photons'' to produce the hard X-ray emission, is found to have a 
$y$-parameter $\sim$ 0.28. The hard X-ray emission illuminates the accretion 
disk and the re-emitted photons produce the observed ``reflection bump''. 
We show that the reflecting medium subtends only a small solid angle  
($\sim 0.15\times 2\pi$), but has a large ionization parameter such that 
iron is ionized up to \ion{Fe}{24}-\ion{Fe}{26}. The presence of a broad iron 
line at $6.58\pm0.04$ keV is also consistent with a highly ionized disk, if we
take into account the gravitational and Doppler shift of the line energy.
These results imply a geometry with a central corona surrounding the black 
hole and the reflection occurring in the innermost region of the disk where 
matter is highly ionized.

\end{abstract}


\section{INTRODUCTION}

Cygnus~X--1 is considered as an archetypical stellar-mass black hole candidate
(BHC). Its observed spectral and temporal X--ray properties have, therefore, 
often been used to distinguish black hole binaries from their neutron star 
counterparts. Though seriously flawed, this approach has resulted in the 
discovery of many BHCs whose candidacy was later confirmed by dynamical mass 
determination based on optical observations. Despite such success, little
is known about the physics behind many observed phenomena in Cyg~X--1, such
as its occasional state transitions and details of its hard X-ray production 
mechanism(s).

Much attention (Belloni et al. 1996; Cui et al. 1997a,b,c; Zhang et al. 1997;
Dotani et al. 1997) has been paid to the recent transition of Cyg~X--1 in
1996 from its
usual hard (or low) state to the rare soft (or high) state discovered by
the All-Sky Monitor (ASM) on {\it Rossi X-ray Timing Explorer} (RXTE) (Cui 
1996), because no similar transitions had been seen since 1980 (Ogawara et al.
1982). This time, nearly continuous coverage of the transition were provided 
simultaneously by two complementary all-sky monitors: ASM/RXTE in the soft 
band (1.3-12 keV) and BATSE/CGRO in the hard band (20-200 keV) through the
entire period. Not only was the anti-correlation (or ``spectral pivoting'') 
between the soft and hard X-rays firmly established, but more surprisingly the
bolometric X-ray luminosity varied little during the transition (Zhang et al. 
1997). This is very different from X-ray outbursts observed in transient BHCs 
during which X-ray luminosity usually changes by more than an order of 
magnitude. 

Snapshots were also taken during the transition and in the soft state with 
more sensitive instruments on several satellites including RXTE (Belloni et 
al. 1996; Cui et al. 1997a,b,c), ASCA (Dotani et al. 1997), and OSSE/CGRO. 
At high energies ($\gtrsim$ 10 keV), the observed spectrum can be represented 
reasonably well by a simple power-law up to $\sim$600 keV in the soft state 
(B.~Phlips 1997, private communication). A steeper power-law is required 
to fit the spectrum at 
low energies ($\lesssim$ 10 keV), in addition to an ultra-soft component (Cui 
et al. 1997a; Dotani et al. 1997). The ultra-soft component is typical of
BHCs (see review by Tanaka \& Lewin 1995, and references therein), and is 
thought to be the emission from the innermost part of the accretion disk. 

The observed spectral and temporal properties of Cyg X-1 both suggest that 
the hard X-rays are likely to be the product of inverse-Compton scattering of 
soft 
photons in the region by energetic electrons in a central corona surrounding 
the black hole (e.g., Ling et al. 1997; Cui et al. 1997c; Dove et al. 1997). 
Possible sources of the seed photons include emission from the accretion disk 
and synchrotron or 
bremsstrahlung emission near the black hole. The hard X-rays emerging from the 
corona may then illuminate the inner disk region, and the re-emitted photons 
from the disk could produce a so-called ``Compton reflection bump'' on the 
observed X-ray spectrum. This spectral feature has been observed for the hard 
state (Done et al. 1992; Ebisawa et al. 1996), but has not for the soft
state, mostly due to the scarcity of high quality data. 

Although ASCA observations are sensitive to soft spectral features
and afford moderate energy resolution ($\Delta E/E \sim$ 8 \% with GIS 
at the iron K-band), the limited  energy range (0.7--10 keV) and 
a small effective area make ASCA insensitive to some spectral 
features such as broad and weak emission lines and/or absorption edges.
On the other hand, RXTE observations cover a much wider energy range 
with a large effective area, but suffer from the converse problem of lacking 
low-energy sensitivity below $\sim$2 keV and energy resolution for studying
narrow line emission.
Therefore, the combination of the two instruments is powerful in disentangling 
various spectral components, thus provides more precise determination of 
the model parameters. In this letter, we present results from simultaneous
ASCA and RXTE observations of Cyg~X-1 near the end of its 1996 transition from
the hard to soft state (Cui et al. 1997a). 

\section{Observations}

Cyg~X-1 was observed on 30 May 1996 by both {\it ASCA} and {\it RXTE} (Dotani 
et al. 1997; Cui et al. 1997a). The observations were scheduled independently,
and fortuitously there was a $\sim$1 ks overlap between them (UT 08:19:40
--- 08:35:37). When the observations were made, the soft X-ray flux had 
already reached the soft-state level (Cui et al. 1997a), implying that the 
evolution of the accretion disk was perhaps completed. However, systematic 
studies of the spectral
and timing properties during this period and comparison of the BATSE/CGRO and 
ASM/RXTE behavior during the ASM soft outburst suggest that the source had not 
settled down in the soft state yet, but very close (Cui et al. 1997a,b). The
observed X-ray spectrum as well as the temporal properties varied during this
transitional period (Cui et al. 1997a,b). It is therefore important to analyze
the simultaneous data together, in order to compare results from two different 
instruments optimized for different energy ranges.

RXTE carries two co-aligned pointing instruments, the {\it Proportional 
Counter Array} (PCA; Jahoda et al. 1996) and the {\it High Energy X-ray 
Timing Experiment} (HEXTE). Due to the limited exposure time, the HEXTE data, 
which cover an energy range of 15-250 keV, suffer from poor statistics. 
Therefore, we only focus our attention on the PCA observation. The PCA has 
an effective area of about 6500 $cm^2$. It covers a 2-60 keV energy range 
with moderate energy resolution ($\sim 18\%$ at 6 keV).
ASCA also has two sets of detectors aboard, 
the {\it Gas Imaging Spectrometer} (GIS; Ohashi et al. 1996; 
Makishima et al. 1996) and the {\it Solid-state Imaging 
Spectrometer} (SIS; Burke et al. 1994). The SIS suffered from
serious photon pile-up as well as severe telemetry saturation because Cyg~X-1
was very bright during the observation (Dotani et al. 1997). Here we use 
only the GIS data.

\section{Analysis and Results}

The details on how the GIS data were reduced and the spectra extracted can be
found in Dotani et al. (1997). For the PCA observation, the data from all five
detectors are used here. The PCA background is now modeled, instead of using 
the earth-occultation data (Cui et al. 1997a), but the difference is 
negligible. To account for calibration uncertainties, we added 1\% 
systematic error to the energy spectra for both the PCA and GIS.
We simultaneously fitted the individual spectra for five detectors in the PCA 
(with response matrices version 2.1.2) and two GIS spectra (with response 
matrices version 4.0 and XRT responses version 2.0 for ancillary response 
files; In addition, the default ``ARF filter'' was applied; cf. Fukazawa et 
al. 1997). The relative normalization between the GIS and 
PCA was allowed to vary due to the uncertainty in the PCA effective area with 
respect to that of GIS. The normalization parameters in the model are then 
determined with the GIS spectra.

First, we tried to find a simple analytic model which can adequately describe 
the observed spectra. Previous studies had shown that an ultra-soft component 
dominates the spectrum at low energies (Cui et al. 1997a; Dotani et al. 1997).
We modeled this component with a multi-color disk model (MCD; Mitsuda et al.
1984; Makishima et al 1986). The addition of a simple power law failed to 
reproduce the spectrum at high energies. 
A reasonable fit was instead obtained with a broken power-law with 
a very gradual high-energy cutoff. However, a careful examination of the PCA 
residual revealed a line feature at 6--7 keV. We added a broad Gaussian 
component to mimic the emission line, and indeed the fit was significantly 
improved. Table~1 summarizes the results of fits with and without the line.
Note that the spectral flattening above $\sim$10 keV is characteristic of 
Compton reflection.

Next, we refitted the spectrum with physical models available, to 
systematically study the effects of inverse Comptonization and 
Compton reflection on the X-ray spectrum. We experimented with MCD for 
the soft component and a Comptonized blackbody model by Nishimura et al. 
(1986) for the hard component (``compbb'' hereafter).  In the compbb model, 
free parameters are temperature and normalization of the blackbody component
(for seed photons), 
and electron temperature and optical depth of the hot plasma. The fit is 
decent (Case 1 in Table~2), but significant residuals are apparent, as shown 
in Figure~1,
in the 3--25 keV band where the PCA calibration uncertainty is small 
($\lesssim$ 2\% for the total response, except for near the Xenon edge at 
$\sim$4.8 keV; K.~Jahoda 1997, internal memo, currently available at 
http://lheawww.gsfc.nasa.gov/users/keith/pcarmfv2.1.2/pcarmfv2.1.2.html).
Inclusion or omission of the data above 25 keV hardly affects the model 
parameters ($<$ 3\%). We added a Gaussian component to account for the line 
at $\sim$6.6 keV. The broad ``bump'' peaked at $\sim$15 keV is 
perhaps due to reflection, so we added a disk reflection component, 
based on a model by Magdziarz \& Zdziarski (1995) (XSPEC model ``pexriv'') 
which assumes a cutoff power law for the illuminating spectrum. Since the 
``compbb'' spectrum with parameters appropriate for Cyg X-1 can be 
approximated by a cutoff power law with a photon index $\sim$ 2.0 and cutoff 
energy $\sim$100 keV, we used the ``pexriv'' model to calculate {\em only}\/ 
the reflected spectrum with these parameters fixed in the fit. We also
assume that the disk temperature is $10^5$ K and its inclination angle 
30\arcdeg\  (same as in Done et al. 1992 and Ebisawa et al. 1996), and the 
gas in the disk is of solar abundances. The only remaining free parameters 
are the normalization (=solid angle $\Omega$) and ionization parameter $\xi$.
The residuals are now much reduced. Figure~2 shows the 
observed X-ray spectrum and the model (Case 2 in Table~2) with each component 
plotted separately.  
The results imply the presence of a highly ionized disk ($\xi>12000$), 
in contrast to the hard state when $\xi \lesssim 100$ (Done et al. 1992; 
Ebisawa et al. 1996). 

The emission line is found to be strong (with an equivalent width $EW 
\simeq 126$ eV) and broad. We also explored the presence of a weak narrow 
line, similar to that observed in the hard state (Done et al.\ 1992; Marshall 
et al. 1993; Ebisawa et al.\ 1996). We fixed the line width to zero, and the 
flux to $\sim 1.0\times 10^{-3}\mbox{ }photons\mbox{ }cm^{-2}\mbox{ }s^{-1}$ 
(i.e., $EW \simeq 13$ eV, similar to that found in the hard state; 
Ebisawa et al. 1996). The fit is clearly worse (comparing Case 3 to Case 2 
in Table~2), so the possibility of a weak narrow line is ruled out.
To better characterize the intrinsic line width, however, we now focus only 
on the ASCA data, taking advantage of its superior energy resolution ($\sim$
0.5 keV FWHM at 6.6 keV). We fixed the continuum parameters as well as the 
line energy to the best-fit values. 
The fit favors a strong broad line of width $E_{\sigma}=0.6$ keV and 
EW 182 eV. Fixing EW at this value and the line width at 30 eV (much
smaller than the instrument resolution; cf. Ebisawa et al. 1996) resulted in 
a larger $\chi^2$ with the F-value, $(\Delta \chi^2/2)/(\chi^2/598) = 6.6$,
thus such a line is ruled out at a confidence level of greater than 
99.9\%. If EW is allowed
to vary, the fit results in a weaker ($EW \simeq 80$ eV) narrow line, but the 
$\chi^2$ is still large, with the F-value 4.1; this can again be rejected
at the ($>$)95\% confidence level. We therefore conclude the iron emission 
line is broad in the soft state. It is also clear that the line energy is 
higher than in the hard state, which is consistent with a higher ionization 
state of the disk.

Dotani et al. (1997) derived a black hole mass of $12\pm^3_1M_{\odot}$, 
based on the ASCA data alone, applying the general relativistic 
accretion disk (GRAD) model by Hanawa (1989) for the soft component. 
They assumed the hard component to be a simple power-law with an iron 
absorption edge, since ASCA data alone is unable to constrain the hard 
component spectral shape precisely. To investigate how the mass estimation 
is affected by the inclusion of the RXTE data and different modeling of the 
hard component, we replaced the MCD model with the GRAD model in Case 2 
and refitted the data. The inferred black hole mass is now
$12.3^{+0.5}_{-0.3}\mbox{ }M_{\odot}$, in excellent agreement with the
ASCA measurement.

\section{Discussion}

As shown in Tables~1 and 2, the ultra-soft component is not very 
insensitive to the choice of models for the high-energy continuum: the 
temperature is in the range 0.39-0.44 keV, and the radius of the emitting
region 72-93 km. The results confirm the fact that the temperature of the 
inner disk is higher in the soft state than in the hard state for Cyg X-1
(Ebisawa et al. 1996; Cui et al. 1997a; Dotani et al. 1997).
This is consistent with the inner edge of the accretion disk being closer to
the black hole in the soft state, because the bolometric X-ray luminosity 
changes little during the state transition (Zhang et al. 1997). 

In addition to the ultra-soft component, the most probable model also requires
a Comptonized hard component, a reflected component, and an emission line 
at $\sim$6.6 keV. It should be noted that the Comptonized 
blackbody model we adopted does not take into account the high-energy spectral
roll-over, so only applies to the energy range much below the cutoff energy at 
$\sim 3 k T_e$. On the other hand, the observed electron temperature for the
hot plasma is quite high, $kT_e\simeq 38$ keV, so the model is likely valid 
over the entire RXTE energy range. In any case, the Compton $y-$parameter 
$(=(4kT_e/mc^2) \tau)$ should be relatively independent of various assumptions 
we made; the best-fit value is 0.28. It is interesting to point out that the 
measured temperature of Compton seed photons is significantly higher than that
of the ultra-soft component and the radius of the emitting region is smaller. 
These seem to suggest that the seed photons are mostly from a region even 
closer to the black hole. However, this interpretation should be taken with 
caution, because the results depend on the choice of models for seed photons.

The results also show that the reflecting medium is highly ionized and its 
subtended solid angle is small ($\sim 0.15\times 2\pi$). In fact, \ion{Fe}{25}
is the most abundant ion ($\sim$46\%), then \ion{Fe}{26} ($\sim$35\%) and 
\ion{Fe}{24} ($\sim$11\%), based on the best-fit parameters. This ionization 
state is much higher than in the hard state, again consistent with the picture
that the optically thick disk is closer to the black hole in the soft state.
The small solid angle implies a geometry of a central corona surrounding the
black hole and the reflection occurring mostly in the innermost portion of the 
disk where gas is highly ionized. 

A broad iron emission line was observed at $6.58\pm0.04$ keV with an
equivalent width $\sim$ 126 eV.
The fluorescent yields for \ion{Fe}{25}, \ion{Fe}{26}, and \ion{Fe}{24} 
are 0.5, 0.7, and 0.75, respectively, which are much larger than that 
for neutral iron (0.34; Krolik \& Kallman 1987). 
The weighted-average line energy taking account of the ion population and
the fluorescent yield is therefore $\sim$ 6.8 keV. 
The lower observed line energy and its broadness can perhaps be attributed to 
gravitational redshift and Doppler effects which become important if the 
emission line originates in the inner region of the disk. A similar line was 
also detected in GX~339-4, as well as an absorption edge at $\sim$8.8 keV 
(Makishima et al 1986), so such features might be common to BHCs in the soft 
state. In comparison, the iron line observed in the hard state is centered at 
$\sim$6.4 keV (Done et al. 1992; Marshall et al 1993; Ebisawa et al. 1996), 
so is likely due to cold iron. This is consistent with the inner disk edge 
being farther away thus cooler in the hard state. The line is broader and 
stronger in the soft state, even though the solid angle subtended by the 
reflecting medium is smaller. This is perhaps due to the combination of 
higher fluorescence yield for highly ionized iron and smaller photoelectric 
absorption of the ionized medium (cf. Krolik \& Kallman 1987; Matt et al. 
1993; Zycki \& Czerny 1994). 

\acknowledgments
We would like to thank S.~N.~Zhang for many useful discussions, and also Piotr
Zycki on Compton reflection models. WC acknowledges support from NASA through 
Contract NAS5--30612.

\clearpage
\begin{deluxetable}{lcccccccccccc}
\scriptsize
\tablecolumns{13}
\tablewidth{0pc}
\tablecaption{\tablenotemark{a}\mbox{ }Fits to a Model of MCD\tablenotemark{b}, Broken
Power-law with a High Energy Cutoff\tablenotemark{c}, and Line\tablenotemark{d}}
\tablehead{
\colhead{$N_H$\tablenotemark{e}}& \colhead{$T_{in}$} & \colhead{$R_{in}$} & \colhead{$E_c$} & \colhead{$E_{\sigma}$} & \colhead{$N_l$} & \colhead{$\alpha_1$}& \colhead{$E_b$} & \colhead{$\alpha_2$} & \colhead{$N_p$} & \colhead{$E_c$} & \colhead{$E_f$} & \colhead{$\chi^2_{\nu}/dof$} \\
 & (keV) & (km) & (keV) & (keV) & & & (keV) & & & (keV) & (keV) &  }
\startdata
$0.53^{+0.01}_{-0.02}$ & $0.379^{+0.007}_{-0.004}$ & $101^{+3}_{-6}$ & \nodata & \nodata & \nodata & $2.56^{+0.01}_{-0.02}$ & $10.9^{+0.1}_{-0.1}$ & $1.41^{+0.03}_{-0.05}$ & $10.4^{+0.2}_{-0.4}$ & $6.7^{+0.1}_{-0.2}$ & $38^{+2}_{-4}$ & $1.81/981$ \nl
$0.52^{+0.02}_{-0.01}$ & $0.386^{+0.006}_{-0.006}$ & $93^{+4}_{-5}$ & $6.56^{+0.04}_{-0.06}$ & $0.57^{+0.07}_{-0.07}$ & $1.5^{+0.1}_{-0.2}$ & $2.66^{+0.01}_{-0.02}$ & $10.7^{+0.1}_{-0.1}$ & $1.85^{+0.02}_{-0.02}$ & $11.6^{+0.3}_{-0.5}$ & $19^{+2}_{-2}$ & $132^{+16}_{-19}$ & $1.34/978$ \nl
\tablenotetext{a}{Errors shown represent 90\% confidence interval.}
\tablenotetext{b}{$T_{in}$ and $R_{in}$ are the temperature and radius of the 
inner disk, respectively, assuming a distance of 2.5 kpc for Cyg~X-1.}
\tablenotetext{c}{$\alpha_1$ and $\alpha_2$ are power-law photon indices, and
$E_b$ the break energy. $E_c$ is the cutoff energy, and $E_f$ the e-folding 
energy. $N_p$ is photon flux at 1 keV in units of $photons\mbox{ }cm^{-2}\mbox{ }s^{-1}$.}
\tablenotetext{d}{$E_c$ and $E_{\sigma}$ are line energy and width, 
respectively, $N_l$ photon flux in units of $10^{-2}\mbox{ }photons\mbox{ }cm^{-2}\mbox{ }s^{-1}$.}
\tablenotetext{e}{Hydrogen column density along the line of sight, in units of $10^{22}\mbox{ }cm^{-2}$.}
\enddata
\end{deluxetable}

\clearpage

\begin{deluxetable}{llcccccccccccc}
\scriptsize
\tablecolumns{14}
\tablewidth{0pc}
\tablecaption{\tablenotemark{a}\mbox{ }Fits to a Model of MCD, Comptonized Blackbody\tablenotemark{b}, Line, and Reflection\tablenotemark{c}}
\tablehead{
\colhead{Case}& \colhead{$N_H$}& \colhead{$T_{in}$} & \colhead{$R_{in}$} & \colhead{$E_c$} & \colhead{$E_{\sigma}$} & \colhead{$N_l$} & \colhead{$T_b$} &\colhead{$T_e$} & \colhead{$\tau$} & \colhead{$R_b$} & \colhead{$\Omega/2\pi$} & \colhead{$\xi$} & \colhead{$\chi^2_{\nu}/dof$} \\
 & & (keV) & (km) & (keV) & (keV) &  & (keV) & (keV) & & (km) & & & }
\startdata
1&$0.40^{+0.02}_{-0.01}$ & $0.441^{+0.004}_{-0.003}$ & $78^{+3}_{-1}$ & \nodata & \nodata & \nodata & $1.01^{+0.01}_{-0.01}$ & $45.4^{+0.4}_{-0.7}$ & $0.86^{+0.01}_{-0.01}$ & $9.4^{+0.2}_{-0.1}$ & \nodata & \nodata & $2.21/983$ \nl
2&$0.40^{+0.01}_{-0.01}$ & $0.436^{+0.004}_{-0.004}$ & $80^{+2}_{-1}$ & $6.58^{+0.05}_{-0.04}$ & $0.35^{+0.07}_{-0.07}$ & $1.0^{+0.1}_{-0.2}$ & $0.90^{+0.02}_{-0.01}$ & $38.7^{+0.3}_{-0.4}$ & $0.93^{+0.01}_{-0.01}$ & $11.8^{+0.4}_{-0.2}$ & $0.15^{+0.03}_{-0.02}$ & $> 11911$ & $1.47/978$ \nl
3&$0.39^{+0.01}_{-0.02}$ & $0.450^{+0.006}_{-0.003}$ & $74^{+2}_{-2}$ & $6.6^{+0.1}_{-0.1}$ & $0$ (fixed)& $0.1$ (fixed)& $1.00^{+0.03}_{-0.01}$ & $35^{+1}_{-1}$ & $0.89^{+0.01}_{-0.01}$ & $9.3^{+0.1}_{-0.2}$ & $0.40^{+0.05}_{-0.02}$ & $296^{+108}_{-100}$ & $1.62/980$ \nl
\tablenotetext{a}{The same as Table~1, unless otherwise noted.}
\tablenotetext{b}{$T_b$: blackbody temperature; $T_e$: electron temperature 
of the Comptonizing corona; $\tau$: optical depth of the corona; and $R_b$ 
is the radius of the blackbody emitting region, assuming a distance of 2.5 
kpc for Cyg~X-1.}
\tablenotetext{c}{$\Omega$ is solid angle subtended by the reflecting medium,
and $\xi$ ionization parameter. }
\enddata
\end{deluxetable}

\clearpage
\begin{figure}
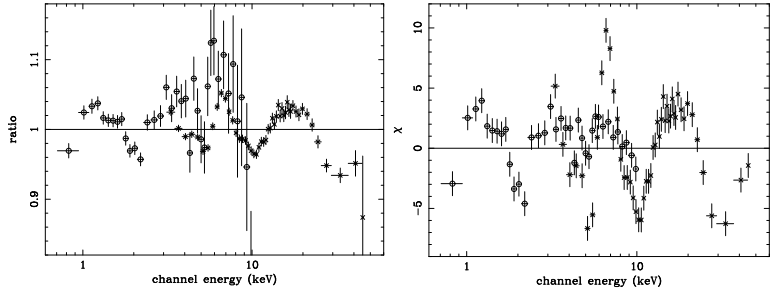

\psfig{figure=fig1a.ps,height=5cm,angle=270}
\vspace{5mm}
\psfig{figure=fig1b.ps,height=5cm,angle=270}
\caption{Residuals from a fit to the observed X-ray spectra with a model of 
MCD and compbb (Case 1 in Table~2). The left panel shows the data-to-model
ratio, while the right one plots the value of $\chi$ to indicate the 
significance of deviations. The GIS data are represented by open 
circles, and the PCA data by asterisks. For display clarity, the spectra 
for two GIS detectors have been rebinned and coadded, so have those for 
five PCA detectors.}
\end{figure}

\begin{figure}
\psfig{figure=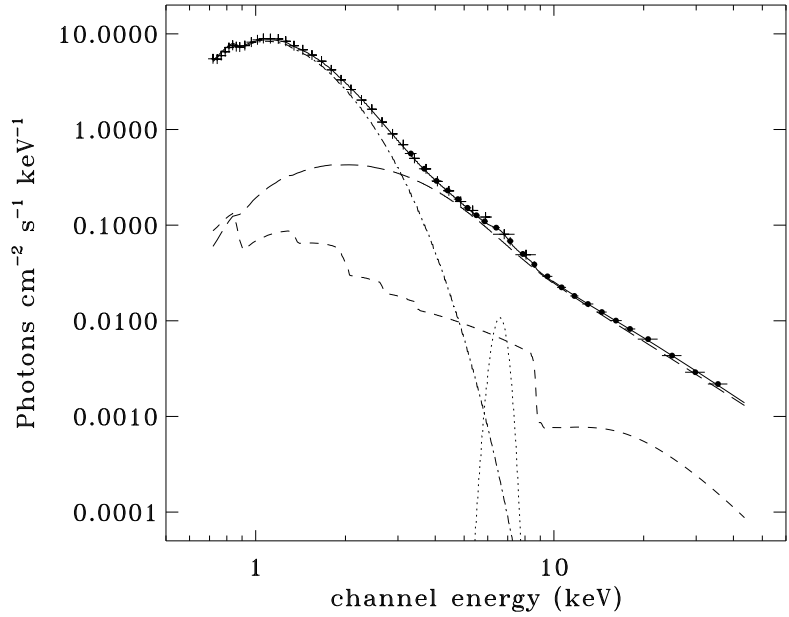,height=9cm}
\caption{Observed GIS (in crosses) and PCA (in filled circles) spectra for 
Cyg X-1, along with the most probable model (Case 2 in Table~2) in solid 
line. The 
individual components are shown separately: MCD in dot-dashed line, compbb 
in long-dashed line, reflection component in short-dashed line, and Gaussian 
in dotted line. For display clarity, the spectra for two GIS detectors have 
been rebinned and coadded, so have those for five PCA detectors. } 
\end{figure}


\clearpage
\begin{thebibliography}{}
\bibitem[Burke et al. 1994]{burke1994}
Burke,~B.~E., Mountain,~R.~W., Daniels,~P.~J., Cooper,~M.~J., \& Dolat,~V.~S. 1994, IEEE trans. nucl. sci. 41, 375

\bibitem[Belloni et al. 1996]{belloni1996}
Belloni,~T., M\'{e}ndez,~M., van der Klis,~M., Hasinger,~G., Lewin,~W.~H.~G., \& van Paradijs,~J. 1996, \apjl, 472, L107

\bibitem[Cui 1996]{cui1996} 
Cui,~W. 1996, \iaucirc\ 6404

\bibitem[Cui et al. 1997a]{cuietal1997a} 
Cui,~W., Heindl,~W.~A., Rothschild,~R.~E., Zhang,~S.~N., Jahoda,~K., \& Focke,~W. 1997a, \apjl, 474, L57

\bibitem[Cui et al. 1997b]{cuietal1997b} 
Cui,~W., Zhang,~S.~N., Heindl,~W.~A., Rothschild,~R.~E., Jahoda,~K., Focke,~W., \& Swank,~J.~H. 1997b, Proceedings of 2nd Integral workshop " The Transparent Universe", Eds: C.~Winkler et al. (St. Malo, France), ESA--SP 382, 209 (astro-ph/9610072)

\bibitem[Cui et al. 1997c]{cuietal1997c}
Cui,~W., Zhang,~S.~N., Focke,~W., and Swank,~J.~H. 1997c, \apj, 484, 383

\bibitem[Done et al. 1992]{done1992}
Done,~C., Mulchaey,~J.~S., Mushotzky,~R.~F., \& Arnaud,~K.~A. 1992, \apj, 395, 275

\bibitem[Dotani et al. 1997]{dotani1997}
Dotani,~T., et al. 1997, \apjl, 485, L87

\bibitem[Dove et al. 1997]{dove1997}
Dove,~J.~B., Wilms,~J., Maisack,~M. \& Begelman,~M.~C. 1997, \apj, 487, 759

\bibitem[ebisawa et al. 1996]{ebisawa1996}
Ebisawa,~K., Ueda,~Y, Inoue,~H, Tanaka,~Y, \& White,~N.~E. 1996, \apj, 467, 419

\bibitem[Hanawa 1989]{hanawa1989}
Hanawa,~T. 1989, \apj, 341, 948

\bibitem[Jahoda et al. 1996]{jahoda1996}
Jahoda,~K., et al. 1996, EUV, X-ray, and Gamma-ray Instrumentation for 
Astronomy VII, O.~H~.W.~Siegmund and M.~A.~Gummin, eds., SPIE~2808, p.~59

\bibitem[Krolik \& Kallman 1987]{krolik1987}
Krolik,~J.~H., \& Kallman,~T.~R. 1987, \apjl, 320, L5

\bibitem[Ling et al. 1997]{ling1997}
Ling,~J., et al. 1997, \apj, 484, 475

\bibitem[Marshall et al. 1993]{marshall1993}
Marshall,~F.~E., Mushotzky,~R.~F., Petre,~R., \& Serlemitsos,~P.~J. 1993, \apj, 419, 301

\bibitem[Magdziarz \& Zdziarski 1995]{magdziarz1995}
Magdziarz,~P., \& Zdziarski,~A. 1995, \mnras, 273, 837

\bibitem[Makishima et al. 1986]{maki1986}
Makishima,~K. et al. 1986, \apj, 308, 635

\bibitem[Makishima et al. 1996]{maki1996}
Makishima,~K. et al. 1996, \pasj, 48, 171

\bibitem[Matt et al. 1993]{matt1993}
Matt,~G., Fabian,~A.~C., \& Ross,~R.~R. 1993, \mnras, 262, 179

\bibitem[Mitsuda et al. 1984]{mitsuda1984}
Mitsuda,~K., et al. 1984, \pasj, 36, 741

\bibitem[Nishimura et al. 1986]{nishimura1986}
Nishimura,~J., Mitsuda,~K., \& Itoh,~M. 1986, \pasj, 38, 819

\bibitem[Ogawara et al. 1982]{ogawara1982}
Ogawara,~Y., et al. 1982, Nature, 295, 675

\bibitem[Ohashi et al. 1996]{ohashi1996}
Ohashi,~T. et al. 1996, \pasj, 48, 157

\bibitem[Tanaka \& Lewin 1995]{tanaka1995} 
Tanaka,~Y., \& Lewin,~W.~H.~G. 1995, in ``X--ray Binaries'', eds. W. H. G. Lewin, J. van Paradijs, \& E. P. J. van den Heuvel (Cambridge U. Press, Cambridge) p. 126

\bibitem[Zhang et~al. 1997]{zhang1997} 
Zhang,~S.~N., Cui,~W., Harmon,~B.~A., Paciesas,~W.~S., Remillard,~R.~E., \& van Paradijs,~J. 1997, \apjl, 477, L95

\bibitem[Zycki \& Czerny 1994]{zycki1994}
Zycki,~P.~T., \& Czerny,~B. 1994, \mnras, 266, 653

\end{thebibliography}
\end{document}